\documentstyle[preprint,prl,aps,epsf]{revtex}

\def\gtwid{\mathrel{\raise.3ex\hbox{$>$\kern-.75em\lower1ex\hbox{$\sim$}}}}
\def\ltwid{\mathrel{\raise.3ex\hbox{$<$\kern-.75em\lower1ex\hbox{$\sim$}}}}
\def\im{{\rm Im}}
\newcommand{\tj}{$t$-$J$\ }

\begin{document}

\draft

\title{The Superconducting Condensation Energy and
       an Antiferromagnetic Exchange Based Pairing Mechanism}
\author{D.J.~Scalapino$^{*}$ and  S.R.~White$^{\dag}$}
\address{
$^{*}$Department of Physics,
University of California,
Santa Barbara, CA 93106
}
\address{
$^{\dag}$Department of Physics,
University of California
Irvine, CA 92697
}
\date{\today}
\maketitle
\begin{abstract}
For the traditional low $T_c$ superconductors, the superconducting
condensation energy is proportional to the change in energy of the ionic
lattice between the normal and superconducting state, providing a clear link
between pairing and the electron--ion interaction.  Here, for the \tj
model, we discuss an analogous relationship between the superconducting
condensation energy and the change in the exchange energy between the
normal and superconducting states.  We point out the possibility of
measuring this using neutron scattering and note that such a measurement,
while certainly difficult,
could provide important evidence for an exchange interaction-based pairing
mechanism.
\end{abstract}
\newpage


During the past several years, a variety of experiments ranging from NMR
\cite{martindale,itoh} and penetration depth \cite{hardy} studies to ARPES \cite{shen,ding} and
Josephson phase interference measurements \cite{wohlman,tsuei} have provided clear
evidence for $d_{x^2-y^2}$-pairing in the high $T_c$ cuprates. This type
of pairing was in fact predicted from a variety of theoretical studies on
Hubbard and \tj models in which a short range Coulomb potential leads 
to a near neighbor exchange interaction and short range antiferromagnetic
correlations \cite{reports}.  Thus, in spite of the differences in the
interpretations of some of these calculations, one might have concluded
that the basic mechanism which is responsible for pairing in the cuprates
arises from the antiferromagnetic exchange interaction and the short range
exchange correlations. However, there is far from a consensus 
on this, and a variety of different basic models and 
pairing mechanisms have been proposed\cite{batlogg}.

In the traditional low temperature superconductors one could see an image
of the phonon density of states $F(\omega)$ in the frequency dependence of the
gap $\Delta(\omega)$ \cite{schrieffer}. One also had a clear isotope effect in some
of the simpler materials and Chester \cite{chester} showed that in this case the
superconducting condensation energy could be related to the change in the
ion kinetic energy. Thus, while the kinetic energy of the electrons is
increased in the superconducting state relative to the normal state, the
decrease in the ion lattice energy is sufficient to give the
condensation energy. This provided a further link between the electron
lattice interaction and the pairing mechanism in the traditional
superconductors. 

Now, in the high $T_c$ cuprates, we believe that one can see the image of
the $k$-dependence of the interaction in $\Delta(k)$ and that this supports
the Hubbard and \tj pictures \cite{scalapino,holestructures}.  
However, as noted, this remains an open
question and it would be useful to look for the analogue of the decrease in
lattice energy and the condensation energy.  From density matrix
renormalization group studies of the \tj model \cite{holestructures}, 
we know that while the kinetic energy of a pair of holes is 
increased relative to having
two separate holes, the exchange energy is reduced. Thus, if the short
range antiferromagnetic spin lattice correlations play 
a similar role to the ion
lattice in the traditional low temperature superconductors, the
condensation energy would be proportional to the change in the exchange
energy between the normal and superconducting states. 

Here we examine this and look for its possible experimental
consequences. Unfortunately, just as in the case of the
traditional electron-phonon systems where the fractional change
in the lattice energy between the normal and superconducting
ground states is small, of order $T_c^2/\mu_F \omega_D$,
and hence hard to detect, here we find that the fractional
change in the exchange energy, of order $T_c^2/\mu_F J$, will
also be difficult to observe. Nevertheless, on a formal level it
is interesting to contrast the relationship between the
superconducting condensation energy and the change in the
exchange energy with a recent proposal by Leggett\cite{leggett} in
which he argues that the condensation energy arises from a
change in the long-wavelength Coulomb energy associated with the
mid-infrared dielectric response.

Our basic idea originated from the results of numerical density
matrix renormalization group calculations\cite{holestructures} for
the \tj model. The \tj\ Hamiltonian in the subspace in which
there are no doubly occupied sites
is given by
\begin{equation}
H = - t \sum_{\langle ij \rangle s}
      ( c_{is}^{\dagger}c_{js}
                + {\rm h.c.}) + 
J \sum_{\langle ij \rangle}
      ( {\bf S}_{i} \! \cdot \! {\bf S}_{j} -
         \frac{n_i n_j}{4} ) .
\label{tj-ham}
\end{equation}
Here $ij$ are near-neighbor sites, $s$ is a spin index, 
$\vec S_i = (c^\dagger_{is}\vec \sigma_{ss'} c_{is'})/2$
and $c^\dagger_{i,s}$ are electron spin and creation operators,
and $n_i= c^\dagger_{i\uparrow}c_{i\uparrow} +
c^\dagger_{i\downarrow}c_{i\downarrow}$.
The near-neighbor hopping and exchange interactions are $t$ and
$J$.  
We have calculated the ground state energy of Eq. (1) for zero
($E_0$), one ($E_1$), and two ($E_2$) holes. For $J/t=0.35$ we
find, for an $8\times8$ system, that the binding 
energy of a pair of holes is 
\begin{equation}
\Delta_B = 2 E_1 - (E_2 + E_0) = 0.23 J .
\end{equation}
We also find that the dominant contribution to this binding
comes from the change in the exchange energy
\begin{equation}
2 \langle J \sum_{\langle ij \rangle} \vec S_i \cdot \vec S_{j} \rangle_1
- \left ( \langle J \sum_{\langle ij \rangle} \vec S_i \cdot \vec S_{j} \rangle_2
+ \langle J \sum_{\langle ij \rangle} 
\vec S_i \cdot \vec S_{j} \rangle_0 \right )
\label{twoa}
\end{equation}
Here 0, 1, and 2 refer to the number of holes in the ground
state. 

The pair binding energy can be used in a simple estimate of
$T_c$: if we relate the superconducting gap to the binding
energy via $2 \Delta = \Delta_B$, and assume that $2 \Delta/kT_c
\approx 6$, we find $T_c \approx 0.04 J/k$. Taking $J=1500 K$,
this gives $T_c \approx 60K$, a quite reasonable value. Now, it is clear
that superconductivity in the cuprates is a much more complicated
phenomena than this simple picture of pair binding. For example,
even in
the $t$-$J$ model, we find that with a finite concentration of
holes, domain walls form, rather than pairs\cite{energetics}. However, the
formation of domain walls in the $t$-$J$ model is also driven
largely by the exchange energy. Therefore, it is reasonable to
assume that whatever the precise mechanism of superconductivity
in the cuprates, energetically it is driven by the exchange
interaction.

Based upon this and in analogy with the electron phonon
case, we suggest that if the basic
interaction which is responsible for pairing in the high $T_c$ cuprates is
the antiferromagnetic exchange, the condensation energy will be proportional
to the change in the exchange energy between the normal and
superconducting phases
\begin{equation}
\frac{\alpha H^2_c(T)\Omega_0}{8\pi} = J\left(\langle\vec S_i
\cdot \vec S_{i+x} + \vec S_i \cdot \vec S_{i+y}\rangle_N 
- \langle\vec S_i \cdot \vec S_{i+x}+ \vec S_i \cdot \vec S_{i+y}
\rangle_S\right)
\label{three}
\end{equation}
Here $H_c(T)$ is the thermodynamic critical field at temperature $T$, 
$\Omega_0$ is the unit cell
volume per $CuO_2$, and $\alpha$ is a factor of order 1. Note
that both expectation values in Eq. (4) are also taken at temperature $T$
with the subscript $N$ referring to a nominal normal state
and $S$ to the superconducting state. Thus one needs to
be able to extrapolate the normal state data to temperatures
$T<T_c$.

For the \tj model we have \cite{hubbard}
\begin{equation}
\left\langle\vec S_i \cdot \vec S_j\right\rangle = 3\int
\ \frac{d^2q}{(2\pi)^2} \int_0^\infty\ \frac{d\omega}{\pi}\ \im\ \chi(q,\omega)
\cos\left[\vec q\cdot (\vec i-\vec j) \right]
\label{one}
\end{equation}
where $\chi(q,\omega)$ is the magnetic susceptibility at temperature $T$. 
For $\vec i$ equal to $\vec j$ we have the sum rule
\begin{equation}
(1-x) S(S+1) = 3 \int\frac{d^2q}{(2\pi)^2}\ \int_0^\infty\ \frac{d\omega}{\pi} 
\ \im\ \chi(q,\omega)
\label{two}
\end{equation}
with $S= 1/2$, and $x$ the hole doping. 
Using Eqs.~(\ref{one}) and (\ref{two}), we can write Eq.~(\ref{three})
in the form
\begin{equation}
\frac{\alpha H_c^2(T)\Omega_0}{8\pi} = 3J\ \int
\ \frac{d^2q}{(2\pi)^2}\ \int_0^\infty\ \frac{d\omega}{\pi}
\ \left(\im\ \chi_S(q,\omega) - \im\ \chi_N(q,\omega)\right)
\left(2-\cos q_x-\cos q_y\right)
\label{four}
\end{equation}
In Eq.~(\ref{four}), we have added a constant 2 using the sum rule
Eq.~(\ref{two}). The form factor $2-\cos q_x
-\cos q_y$ favors large momentum transfers $q_x \sim q_y \sim \pi$ and the
energy scale is set by $\omega \ltwid J$. 

For the optimally or possibly the overdoped materials, it may be
that $\im\ \chi_N(q,\omega)$ has reached its ``low temperature
normal form'' at temperatures above $T_c$. In this case, one could
extract it from neutron scattering data for $T>T_c$. Then, using
low temperature $T<<T_c$ data for $\im\ \chi_S(q,\omega)$ in Eq. (\ref{four}), 
one would obtain the condensation energy. 
Because
$H^2_c(0)\Omega_0/8\pi J\sim 10^{-3}$, it will require
extremely careful neutron scattering measurements to check
Eq.~(\ref{four}). 
Furthermore, one will have to be satisfied that the
normal state measurements taken at temperatures above $T_c$ can be
extrapolated to a temperature which is
low compared to $T_c$. Clearly, this will be difficult.  
However, on a
formal level, it is interesting to contrast the content of
Eq.~(\ref{four}) with the recent proposal by Leggett \cite{leggett}. He takes
the point of view that the pairing mechanism is associated with the long
wave length Coulomb energy
and relates the condensation energy to a change in the
dielectric function between the normal and superconducting state. He then
argues that the important contributions are associated with momentum
transfers which are small compared to $\pi$ and energy transfers in the
mid-infrared region, 0.1 to 1.5eV.

Now, it is certainly true that if one goes all the way back, the Coulomb
energy is responsible for the exchange interaction we have focused on. However,
having integrated out the short range part of the Coulomb interaction to
arrive at an exchange interaction
$J\sim 4t^2/U$, we conclude from Eq.~(\ref{four})
that the important part of the pairing
interaction is associated with large momentum transfers $q\sim(\pi/a,
\pi/a)$ and energies less than or of order $J\sim 0.1$eV.
Thus, contrary to ref.~\cite{leggett}, where one seeks to find a relationship
between the condensation energy and the change in the {\it dielectric}
 response
between the normal and superconducting state in the small momentum and
higher energy
0.1--1.5eV regime, we suggest that the condensation energy is related to 
changes in the {\it magnetic} spin response at large momentum transfer and 
energies $\omega\ltwid J$.

Thus, it would be very interesting if it were possible to confirm or
contradict the relationship of the change in $\langle \vec S_i \cdot
\vec S_{i+\hat x}\rangle$ between the normal and superconducting states and
the superconducting condensation energy given by Eqs.~(\ref{three})and 
(\ref{four}). 
\vskip.50in
\centerline{ACKNOWLEDGMENTS}

We thank A.J.~Leggett and S.C. Zhang for interesting discussions. DJS
acknowledges support from the Department of Energy under grant,
DE-FG03-85ER-45197 and SRW acknowledges NSF suport under grant DMR95-09945.
DJS would also like to acknowledge the Program on Correlated
Electrons at the Center for Material Science at Los Alamos
National Laboratory.


\begin{thebibliography}{10}
\bibitem{hubbard} For the Hubbard model the left hand side of
Eq. (6) is replaced by $\langle \vec S^2\rangle$, which could
change slightly in the superconducting state, so that one would
not want to make the subtraction in Eq. (7) and the 2 in Eq. (7) should be
dropped.
\end{thebibliography}
\end{document}